\documentclass{new_tlp}
\usepackage{mathptmx}
\usepackage{listings}
\usepackage{textcomp}

\newcommand\bcmdtab{\noindent\bgroup\tabcolsep=0pt%
  \begin{tabular}{@{}p{10pc}@{}p{20pc}@{}}}
\newcommand\ecmdtab{\end{tabular}\egroup}

\newcommand{\myRef}[2]{\textbf{#1~\ref{#2}}} 
\usepackage{graphicx}
\newcommand{\myFig}[4]{
\begin{figure}
\centering
\includegraphics[width=#4\linewidth,height=\textheight,keepaspectratio]{assets/#1}
\caption{#2}\label{#3}
\end{figure}
}

  \title[Refactoring Whitby for Clean Architecture]
		{Refactoring the Whitby Intelligent Tutoring System\\ for Clean Architecture}

  \author[P. S. Brown et al.]
		 {Paul S. Brown, Vania Dimitrova\\
		 University of Leeds, Leeds, UK\\
		 \email{sc16pb@leeds.ac.uk ; v.g.dimitrova@leeds.ac.uk}
		 \and Glen Hart\\
			Defence Science and Technology Laboratory [dstl]
		\and Anthony G. Cohn\\
		University of Leeds; Qingdao University of Science and Technology; Tongji University; Shandong University\\
		\email{a.g.cohn@leeds.ac.uk}
		\and Paulo Moura\\
			Center for Research in Advanced Computing Systems, INESC-TEC, Portugal\\
			\email{pmoura@logtalk.org}}

\pagerange{\pageref{firstpage}--\pageref{lastpage}}

\begin{document}

\label{firstpage}

\maketitle

\begin{abstract}
Whitby is the server-side of an Intelligent Tutoring System application for
learning System-Theoretic Process Analysis (STPA), a methodology used to ensure
the safety of anything that can be represented with a systems model. The
underlying logic driving the reasoning behind Whitby is Situation Calculus,
which is a many-sorted logic with situation, action, and object sorts. The
Situation Calculus is applied to Ontology Authoring and Contingent Scaffolding:
the primary activities within Whitby. Thus many fluents and actions are
aggregated in Whitby from these two sub-applications and from Whitby itself,
but all are available through a common situation query interface that does not
depend upon any of the fluents or actions. Each STPA project in Whitby is a
single situation term, which is queried for fluents that include the ontology,
and to determine what pedagogical interventions to offer.

Initially Whitby was written in Prolog using a module system. In the interest
of a cleaner architecture and implementation with improved code reuse and
extensibility, the
initial application was refactored into Logtalk. This refactoring includes
decoupling the Situation Calculus reasoner, Ontology Authoring framework,
and Contingent Scaffolding framework into third-party libraries that can be
reused in other applications. This extraction was achieved by inverting
dependencies via Logtalk \textit{protocols} and \textit{categories}, which are
reusable interfaces and components that provide functionally cohesive sets of
predicate \textit{declarations} and predicate \textit{definitions}. In this
paper the architectures of two iterations of Whitby are evaluated with
respect to the motivations behind the refactor: clean architecture enabling
code reuse and extensibility. 

\textbf{UNDER CONSIDERATION FOR ACCEPTANCE IN TPLP}.
\end{abstract}

  \begin{keywords}
	Architecture, Dependency Inversion, Prolog Modules, Logtalk
  \end{keywords}


\section{Introduction}

System-Theoretic Process Analysis (STPA),
is an emerging methodology used by system safety analysts from an initial
conceptualisation before the system design, through to a loss occurring. Trying
to determine how a hypothetical system should be designed, built, and
maintained in order to prevent potentially catastrophic losses is a
difficult and cognitively demanding task. To aid with this task an application
has been developed and deployed for a select group in
order to test the efficacy of the pedagogical techniques employed within the
application. The server-side part of this application is called ``Whitby''.

Within Whitby three primary domains are discussed. Although it is not necessary
to understand these domains to consider the architecture, they are introduced
here for orientation:
\begin{itemize}
	\item Situation Calculus: a many-sorted second order logic for reasoning about situations and actions. The definition used is that of \citeN{reiter:2001}.
	\item Ontology Authoring: the knowledge engineering process of defining a formal, ontological model of some simplified world for a purpose~\cite{gruber:1995}.
	\item Contingent Scaffolding: a pedagogical technique in which immediate intervention is offered to a struggling learner at a level of intrusion based upon previous behaviour~\cite{wood:1976}.
\end{itemize}

Prolog was chosen as an implementation language from the outset due to using
both Ontology and Situation Calculus as foundations in the design. However, the
entire application has been re-written twice to overcome the architectural
issues that forced unsatisfactory solutions. The ramifications of these compromises grew with the complexity
of the application. Some of these issues are discussed in \myRef{Section}{sec:whitbyarch}.

In the final rewrite the code base was transitioned to Logtalk and sufficiently decoupled
as to extract three libraries. This extraction was the main motivation behind the rewrites:
the prior version was functioning but useful parts of it couldn't be shared.
The application, including these three libraries, 
contains approximately 10,000 lines of Logtalk source code plus the GUI written in
ClojureScript. The parts authored in Logtalk cover persistence of user
projects, situation calculus reasoning, ontology authoring, contingent
scaffolding, web server, and generating HTML including forms based upon the
history of a user project.

In this paper, the software engineering principles that guided the application rewriting
are presented, the architectures of the logic programming parts of the last two
versions of Whitby are compared, the motivations behind the transition from
Prolog to Logtalk are discussed, and the lessons learned are summarized.

\section{The Dependency Inversion Principle}

In refactoring the architecture, the \textit{SOLID principles} of clean
architecture~\cite{martin:2018} are applied. These principles are the summary
of 20 years of debate between developers attempting to abstract what made their
software maintainable and extensible. They are intended to avoid the situation,
observed in even market-leading software, where progress is slowing while
cost per line of code increases, all while increasing development
staff~\cite{martin:2018}. 

Appropriate application of these principles produces code that is easy to read,
maintain, extend and test. These benefits are primarily to those developing and maintaining
the software, which then has implications to organisations producing and consuming
the software over a period of time. Software with a clean architecture is argued to
be easily extensible with new features, typically using a plugin architecture,
and robust to changes in business rules, technology, and deployment
scenarios~\cite{martin:2018}. Thus it reduces application risks
and development costs.

The principles of SOLID architecture are:
\begin{itemize}
	\item \textbf{S}ingle Responsibility Principle: each part has one and only one
	reason to change; it is accountable to one stakeholder
	\item \textbf{O}pen-Closed Principle: code should be open to extension and closed
	for modification; such as in a plugin architecture where new features are created
	by adding new code rather than editing existing code
	\item \textbf{L}iskov Substitution Principle: parts should be interchangeable, which
	makes it robust to even significant changes such as to business rules meeting new 
	legal requirements, or to swapping components such as the database used or GUI framework
	\item \textbf{I}nterface Segregation Principle: do not depend on things not used, which
	makes dependents of some part robust against changes required by other dependents
	of that part
	\item \textbf{D}ependency Inversion Principle: high-level policy should not depend
	on low-level details, but details should depend on policies, such that code which
	is volatile is not depended upon by code that is stable
\end{itemize}

\myFig{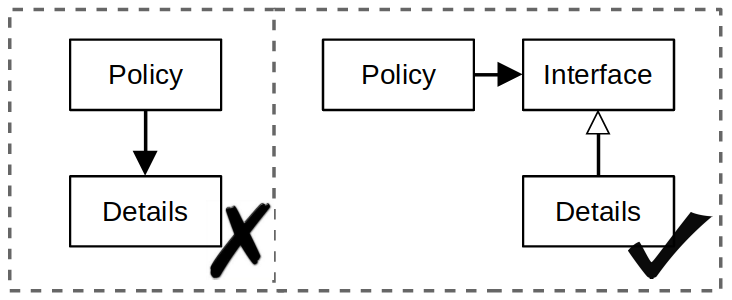}{Dependency Inversion Principle. Left-hand side has high-level policy depending on
low-level details, which is not recommended. Right-hand side has the dependency inverted by the policy
depending on some interface, which the details extend.}{fig:dip}{0.65}

The Dependency Inversion Principle, depicted in \myRef{Fig}{fig:dip}, is the main principle driving
this refactor and
the technique used for decoupling. Closely related to
this principle is the
\textit{code for interface, not for implementation} best practice: no concrete
module should be imported into any other. Instead, an \textit{abstract definition}
of what the module should provide is used. This principle and best practice allows
high-level policies to be left untouched as low-level details are swapped or undergo
change, which in turn makes reuse of the high-level principles as libraries possible.
The concept of interface is thus central to the application of this principle and
best practice as further discussed in \myRef{Section}{sec:dipwlgt}. In Logtalk,
interfaces are represented using \textit{protocols}, but Prolog module systems
do not provide an equivalent feature\footnote{The ISO Prolog standard for
modules~\cite{isoiec00} does specify a module interface language construct but
only allows a single implementation per interface, thus defeating the main purpose
of defining interfaces. Moreover, this standard is ignored by Prolog systems.}.

The dependence on
an interface, shown in \myRef{Fig}{fig:dip}, also differentiates the technique from
dependency injection or meta-programming where the context of
the low-level details are passed to the high-level policy. With or without meta-programming,
the policy is dependent on the predicates required being present in the details,
however with meta-programming the interface is implicit, ungoverned, and not self-documenting.
By using a declared interface as a first-class language feature the required predicates
become explicit, the details are governed through a declared promise to define the interface, implementers of the interface can be enumerated using language reflection predicates, and documentation can be automatically generated. 

These principles are also considered at the component level~\cite{martin:2018}.
At this level of abstraction, the key ideas are to enable \textit{reuse} through
sensible component contents and \textit{decoupling} through dependency cycle
elimination as well as correlating dependency with stability.

\section{Whitby before refactoring: OWLSAI}
\label{sec:whitbyarch}

\myFig{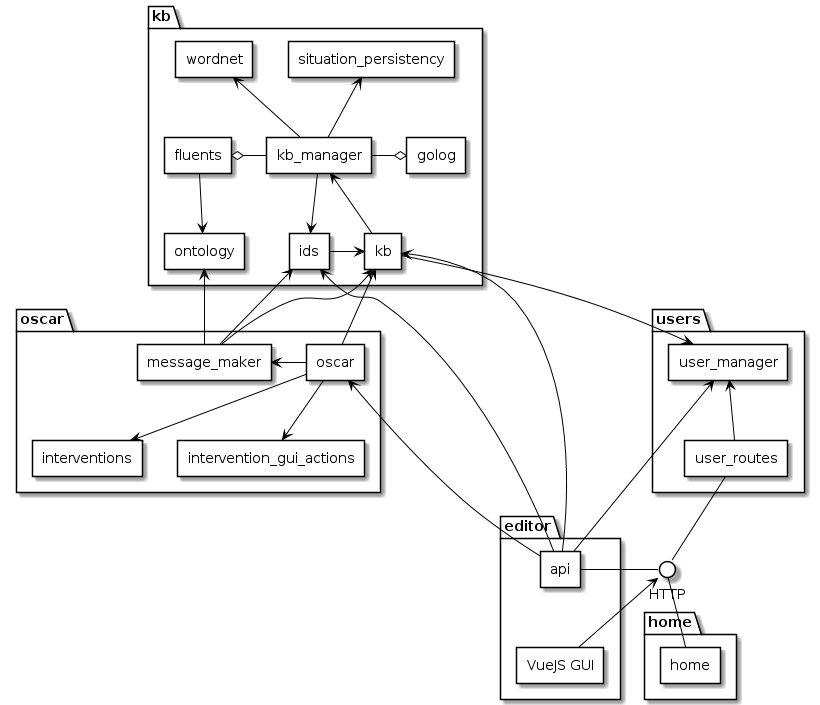}{Dependencies in OWLSAI. Each node is a file, within their
directories, which distinguish modules. Arrows denote imports, open diamonds
denote consults.}{fig:owlsai}{1}

The first version of Whitby, written in Prolog using the module system and
depicted in \myRef{Fig}{fig:owlsai}, was originally called \textit{OWLSAI} (Web
Ontology Language Safety Artificial Intelligence). This application did
implement the features required of it\footnote{The only feature requirement that changed
between OWLSAI and Whitby was a change from handling users with logins and many
projects, to only handling many projects. Therefore this aspect is not compared.}.
There was no particular issue with it from a \textit{user} perspective. The issues
were entirely at the \textit{developer} experience level.

The \texttt{kb} directory in \myRef{Fig}{fig:owlsai} is responsible for the
ontology authoring with situation calculus. The \texttt{oscar} directory, which
is responsible for the contingent scaffolding interventions, is only dependent
upon the code within this module. This dependence is a necessity as
\texttt{oscar} needs to know about a user's ontology in order to offer relevant
interventions.

Several violations of the SOLID principles are hidden at this level of
abstraction. Note that \texttt{golog} and \texttt{fluents} are aggregated in
\texttt{kb\_manager}. This compromise is necessary because
the Golog Situation Calculus reasoner from \citeN{reiter:2001} includes a
predicate that calls these fluents, and so \texttt{golog} is dependent upon
\texttt{fluents}. It is not uncommon in Prolog to apply some set of rules, like
those in \texttt{golog}, to some facts, like those in \texttt{fluents}.

But for \texttt{golog} to reason over the \texttt{fluents} and
\texttt{actions} defined 
for some particular world under analysis requires that world (details) to
be imported into \texttt{golog} (policy). In this manner, the abstract is dependent
upon the concrete, the stable is dependent upon the flexible, the calculus is
dependent upon its own application. It's a violation of clean architecture
that prevents code reuse: \texttt{golog} cannot be extended to include a
defined world to reason over without modification to its own source code.
Concurrent handling of multiple defined
worlds is also precluded.
Although there are workarounds to these problems, some of which are
discussed here, they are unsatisfactory as the lack of necessary
language constructs to cleanly express the
application architecture results in the violation of SOLID principles.

To circumvent the issue of circular dependencies in OWLSAI, the \texttt{golog}
and \texttt{fluents} files were consulted instead. This loaded them both into the \texttt{kb\_manager} namespace. 
However, this causes a conundrum to resolve as there are
fluents and actions for the \texttt{oscar} module that need to be defined in the
\texttt{fluents} and \texttt{kb\_manager} file so that \texttt{golog} is in the
same namespace as them. For example, actions pertaining to ontology authoring,
contingent scaffolding, and user interface are defined adjacent to each other
in \texttt{kb\_manager}, in violation of the Single Responsibility Principle, as
seen in this snippet:

\begin{lstlisting}
:- consult(kb(golog)).
:- consult(kb(fluents)).

%! action(Action, GologPossQuery)

% Ontology Authoring
action( add_data(_User, _Time, Payload),
        -asserted(Payload)).
action( delete_data(_User, _Time, Payload),
        asserted(Payload)).
        
% User Scaffolding Actions
action( dismiss_intervention(_User, _Time, Fact, Level),
        intervened(_, Fact, Level, _)).
action( request_intervention_increase(_User, _T, ID, Fact, Level),
        intervened(ID, Fact, Level, _)).

% Agent Scaffolding Actions
action( intervene(_User, _Time, Fact, Level, _Payload),
        -some(n, (dismissed(Fact, n) & n >= L))).
        
% User UI Actions
action( navigate_to_step(_User, _Time, _Step), true).
action( concept_focus(_User, _Time, _Focus), true).
action( glossary_lookup(_User, _Time, term(_Term)), true).
action( nudge(_User, _Time, _R), true).
\end{lstlisting}

Code belonging to \texttt{oscar} resides in files in
\texttt{kb} that is loaded into a different module. It should reside in files
in \texttt{oscar} that are somehow made visible to \texttt{kb} to
maintain separation of responsibilities and to ease code navigation. There are
mechanisms to achieve this in Prolog: via consulting which would warn if a predicate
were redefined, or via \texttt{include/1}, which includes the text of the file within
the other.

Another mechanism tried for including actions from different modules into \texttt{kb\_manager}
was to declare \texttt{action/2} as a multifile predicate in \texttt{kb\_manager}. Clauses for
the predicate could then be defined in \texttt{oscar} and any other module by
using a prefix: \texttt{kb\_manager:action(\dots)}. However, this violates the Dependency
Inversion Principle as high-level policy predicates belonging to general ontology
authoring and scaffolding are then dependent on the low-level detail that is the
\texttt{kb\_manager}, which is the module responsible for updating and querying user projects extending the ontology. Furthermore, this prefix referring to a specific, fixed module would
prohibit the substitution of that module, thus violating the Liskov Substitution Principle of
SOLID. 

To use \texttt{include/1} directives or multifile predicates would be to
take code from \texttt{oscar} and
have it effect the behaviour of \texttt{kb}; thus a developer working on either
module must understand how the other one is working. A poorly placed cut,
unfortunately named predicate, or redefinition of an operator in \texttt{oscar}
could cause \texttt{kb\_manager}, upon which it depends, to no-longer function
correctly. It opens up the potential for the consumer of some code to break
what it should only depend upon. A developer debugging \texttt{kb}
looking solely at their code in \texttt{kb}, believing it has no dependencies as the
architecture diagram shows, would have little hope of resolving such an error. For these 
reasons module systems are favoured over the older \texttt{consult/1} and \texttt{include/1}
predicates and why using them is also an unsatisfactory solution. When authoring OWLSAI,
the more robust unsatisfactory solution was chosen, putting \texttt{oscar} code into \texttt{kb},
violating the Open-Closed principle and preventing code reuse, but easing debugging.

Ideally a module would be used but the dependency needs to be inverted, such that \texttt{fluents}
depend upon \texttt{golog}.
One workaround within the module system would be to make \texttt{kb\_manager} dependent upon
\texttt{fluents} and \texttt{golog} directly. Then the module can be included with the
fluent or action where it is defined in all queries to
Golog\footnote{Available at: http://www.cs.toronto.edu/cogrobo/kia/}. 
For example:
\begin{lstlisting}
holds(Module:Fluent0, Situation) :-
  Module:restoreSitArg(Fluent0, Situation, Fluent),
  Module:Fluent.
holds(Module:Fact, Situation) :-
  not Module:restoreSitArg(Fact, Situation, _), 
  isAtom(Fact),
  Module:Fact.
\end{lstlisting}

This example also represents the resultant code of one strategy attempted via
using meta-predicates to invert the dependency without using the interface
depicted in \myRef{Fig}{fig:dip}. With \texttt{holds/2} defined as a
meta-predicate the calling context is passed implicitly, but
\texttt{restoreSitArg/3} will be defined in the same module as the \texttt{Fluent},
which may be in a different module from the calling context: in Whitby there
are multiple calling contexts, whereas each fluent is defined once. To make
\texttt{restoreSitArg/3} available to the calling context would result in
name-clashes when more than one module defining fluents is used. Therefore the
dependency module where the definitions reside needed to be passed (or injected)
for context as per this example.

The concern for this example is in the Golog call to 
\texttt{Module:Fluent}, where \texttt{Fluent} could be anything, including a meta-predicate, given in the query, which is a qualified call potentially breaking the encapsulation of the module.
Furthermore, it's no longer possible to use \texttt{holds/2} with a variable as the first
argument to find fluents that hold in a ground situation without explicitly enumerating
all modules and testing if they define fluents or not; the lack of protocols/interfaces
as first-class entities precludes a simple and clean enumeration of only those modules
that would declare conformance to a given protocol. 
The import semantics of Prolog modules also would force the use of these
explicitly-qualified calls for the conforming modules to prevent predicate import
clashes. This goes against what is considered best practice with Prolog modules:
the use of implicit imports and implicit module-qualified predicate calls. But that
is not the primary issue: by making a module that defines fluents and actions
an \textit{explicit} argument, we are forced to anticipate all predicates that,
although not accessing fluents and actions directly, may be indirectly calling a
predicate that requires that access (and thus require the module argument to be passed
from upstream). 

In Whitby however, which makes use of the required
language constructs provided by Logtalk, \texttt{SitCalc} is loaded as a third-party
package. As Logtalk does not use module-like imports semantics, there are never any
loading conflicts when two or more loaded objects define the same public predicates.
Furthermore, Whitby also loads packages defining ontology authoring terms and
contingent scaffolding terms. The only place in Whitby where the contents of those
packages need be considered is in the use of their fluents in queries of a situation 
and in the doing of their provided actions, both of which are done without the
requirement to explicitly define the correct context to reason about them in.

Although dependency inversion is the crucial issue, there are additional
violations of clean architecture that need to be addressed. The dependency
cycle between \texttt{kb\_manager}, \texttt{ids}, and \texttt{kb} can cause a
small edit in one of them to have perpetual ramifications as its dependency
graph is also adapted to the change. Golog is more than a Situation Calculus
reasoner; it is a parser for a Situation Calculus based language;  thus
\texttt{kb} is depending on code that it does not use. Furthermore, the four
dependencies from the \texttt{oscar} module to the \texttt{kb} module suggest
substitution would require more effort than necessary.

\section{Refactored Whitby}

\myFig{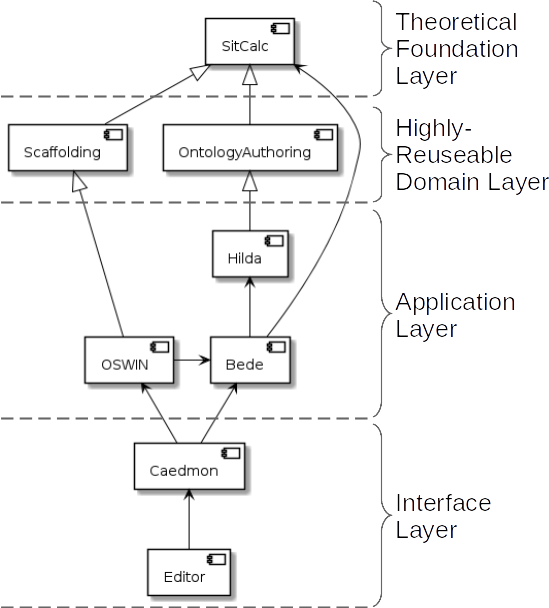}{Dependencies between components of Whitby. Open arrows denote extension, closed arrows denote dependence.}{fig:whitbyabstract}{0.75}

\myFig{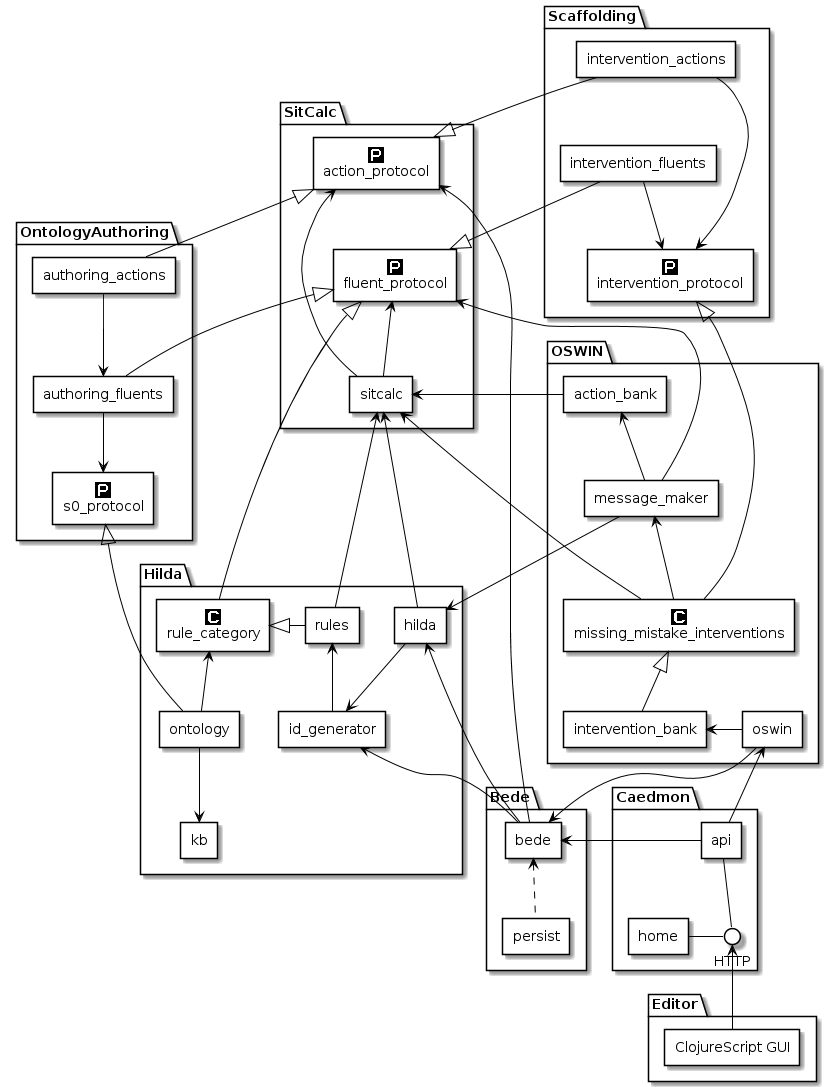}{Dependencies in Whitby and extracted libraries. Each node (without a mark) is an object, within their
directories. Protocols are marked with a ``P'', categories with a ``C''. Closed arrows denote dependence, open arrows denote implementation or extension, dashed arrow denotes event monitoring.}{fig:whitby}{1}

The abstract architecture of Whitby is depicted in \myRef{Fig}{fig:whitbyabstract}, whereas a detailed view is in \myRef{Fig}{fig:whitby}. From the abstract view it can be seen how
Whitby was designed to decouple the components of OWLSAI enabling code reuse. It is not possible
to layer OWLSAI in a similar manner due to the compromises made and tight-coupling.

\texttt{SitCalc} provides the theoretical foundation,
which can be used to tackle a multitude of problems, it depends on nothing. The next
``Highly Reusable Domain Layer'' is the application of \texttt{SitCalc} to two domains;
these libraries depend on \texttt{SitCalc}, but nothing in Whitby. Therefore they
can be reused by any application wishing to apply Situation Calculus to Contingent Scaffolding
or Ontology Authoring. The ``Application Layer'' is the core of the Whitby application, it is
this code that applies the reusable libraries to the particular task
at hand: Contingent Scaffolding an STPA analyst who is unwittingly authoring an extension
to an ontology. Finally the ``Interface Layer'' provides a convenient means for the user to
interact with the application.

Whitby loads \texttt{SitCalc}, \texttt{OntologyAuthoring}, and
\texttt{Scaffolding} defined as third-party libraries. In this application a
naming convention around Whitby Abbey was adopted to aid in organising the
code, thus the modules are:
\begin{itemize}
	\item \textbf{OSWIN}: \textbf{O}ntology-driven \textbf{S}caffolding
	\textbf{W}ith \textbf{I}nteractive \textbf{N}udges (extends \texttt{Scaffolding})
	\item \textbf{Hilda}: The wise, Hilda handles the ontology authoring (extends \texttt{OntologyAuthoring})
	\item \textbf{Bede}: The historian, records the actions that are done
	\item \textbf{Caedmon}: The poet, responsible for the user interface
\end{itemize}

\myFig{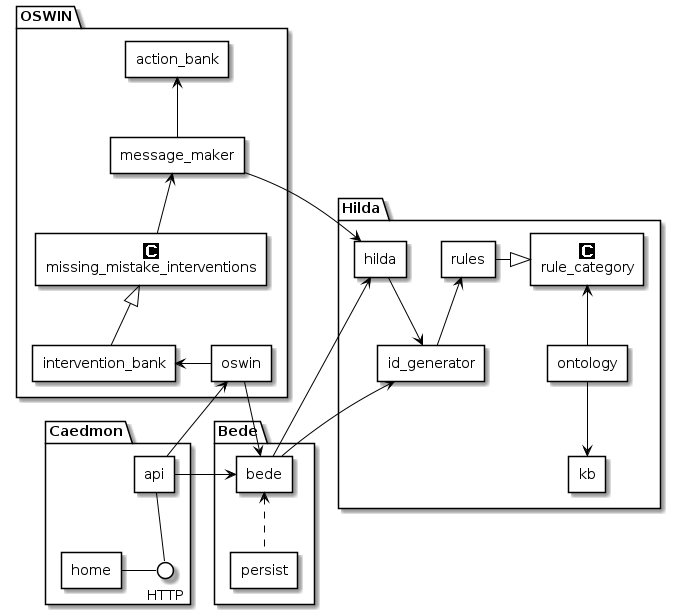}{Dependencies within Whitby only. Each node is an object, within
their directories. Arrows denote dependence, open arrows denote implementation
or extension, dashed arrow denotes event monitoring. Categories are marked with a ``C''.}{fig:whitbyonly}{0.85}

The architecture, shown in \myRef{Fig}{fig:whitby}, initially appears more complex
than OWLSAI in \myRef{Fig}{fig:owlsai} as the third-party libraries that were extracted
are also included, together with the protocols used to achieve dependency inversion.
\myRef{Fig}{fig:whitbyonly} shows only the internal dependencies: how the
application appears to a developer working on it. Such a developer need not
concern themselves with the working of any of the imported libraries; they are
only responsible for what is depicted in \myRef{Fig}{fig:whitbyonly}. For
example, Whitby required a fluent describing what the user is looking at in the GUI;
this is particular to the application of Whitby and so is not defined in
\texttt{OntologyAuthoring}. To add this fluent to Whitby requires creating a
new object that conforms to the \texttt{fluent\_protocol}: new behaviour via
extension rather than modification and exposing fluents that the application
developer has no business editing.

\myRef{Fig}{fig:whitbyonly} is a cleaner architecture, with no dependency
cycles. However it is not yet perfect. For example, \texttt{bede} should not
depend upon \texttt{id\_generator}. That particular predicate should be exposed
through \texttt{hilda}, which provides an interface enabling easier
substitution of the objects that \texttt{hilda} depends upon. Early in the refactoring
to Whitby, each of the named directories was implemented as its own microservice
communicating over HTTP. Correcting this issue would make it simple to split Whitby back into
microservices for scalability, which isn't possible to achieve with OWLSAI due
to the tight coupling between components.

\section{Dependency Inversion using Logtalk Protocols}
\label{sec:dipwlgt}

To achieve the desired architecture requires the application of the Dependency
Inversion Principle, which can be accomplished via the Abstract Factory
design pattern~\cite{gof:1997,martin:2018} described as:

\begin{quote}
    Provide an interface for creating families of related or dependent objects
    without specifying their concrete classes.~\cite{gof:1997}
\end{quote}

In logic programming, we can reinterpret the implicitly imperative idea of 
\textit{creating families} as \textit{declaratively defining families}. Therefore,
with Logtalk it becomes possible to do Dependency Inversion without dynamically creating
objects. The concept of interface, in turn,
is readily available using Logtalk \textit{protocols}, as described below.

The Dependency Inversion Principle is applied to decouple the application into three major
components. First a \texttt{SitCalc} library is extracted. Then \texttt{SitCalc} is extended, not modified, to
create \texttt{OntologyAuthoring} and \texttt{Scaffolding} libraries. Finally,
Whitby is created by importing these libraries as 
third-party libraries. The final architecture, with these libraries
included, is shown in \myRef{Fig}{fig:whitby}. We start with a brief overview of Logtalk
followed by a detailed account of how we applied this design principle to each component.

\subsection{Logtalk Overview}

Logtalk as a language reinterprets object-oriented concepts from first principles to provide
logic programming with code encapsulation and code reuse mechanisms that are key in expressing
well understood design principles and patterns (described in depth in
``The Logtalk Handbook'' \cite{TheLogtalkHandbook}). A key feature is the clear distinction between \textit{predicate declarations} and
\textit{predicate definitions}\footnote{This distinction exists in standard
Prolog~\cite{isoiec95} only for predicates declared as \textit{dynamic} or \textit{multifile}.
Notably, \textit{static} predicates exported by a module must be defined by the module.},
which can be encapsulated and reused as follows:
\begin{itemize}
    \setlength\itemsep{1em}
    \item \textit{protocols}: Group \textit{functionally cohesive} predicate declarations that can
    then be implemented by any number of \textit{objects} and \textit{categories}. Allows an object
    or category to promise conformance to an interface.
    
	\item \textit{objects}: Group predicate declarations and predicate definitions. Objects can be
	\textit{stand-alone} or part of hierarchies. Object enforce encapsulation, preventing calling predicates that are not within scope. Predicates are called
	using \textit{message sending}, which decouples calling a predicate from the predicate definition that is used to answer the message.
	
	\item \textit{categories}: Group a \textit{functionally cohesive} set of predicate declarations
	and predicate definitions, providing a fine-grained unit of code reuse that can be imported by
	any number of objects, thus providing a \textit{composition} mechanism as an alternative to the use	of inheritance.
\end{itemize}

Predicates can be declared \textit{public}, \textit{protected}, or \textit{private}.
A predicate declaration does not require that the predicate is also defined.
Being able to \textit{declare} a predicate, independent of any other predicate properties,
without necessarily \textit{defining} it is a fundamental requirement for the definition of
protocols. It also provides clear \textit{closed world semantics} where calling a
declared predicate that is not defined simply fails instead of generating an error
(orthogonal to the predicate being \textit{static} or \textit{dynamic}).

Logtalk defines a comprehensive set of \textit{reflection} predicates for reasoning
about the use of these components in the program. In particular, the \texttt{conforms\_to\_protocol/2}, which is true if the first argument implements or is an extension of something that implements the protocol named in the second argument, and  \texttt{current\_object/1}, which is true if its argument is an object in the application current state (categories and protocols have their own counterparts). These predicates are
used in the implementation of the SOLID principles as illustrated in the next sections.

Logtalk also provides a comprehensive set of portable developer tools, notably for
documenting, diagramming, and testing that were used extensively. These tools reflect
how the language constructs are used in applications, from API documentation to
diagrams at multiple levels of abstraction that help developers and maintainers
navigate and understand the code base and its architecture.

\subsection{A Reusable SitCalc Library}

The \texttt{SitCalc} library includes predicates that need to send messages to
fluent and action objects. Rather than depend on these fluents and actions directly,
it depends instead on objects conforming to \texttt{action\_protocol} and
\texttt{fluent\_protocol}. This is the Dependency Inversion Principle of SOLID:
to depend only on protocols/interfaces and not on concrete code~\cite{martin:2018}.
These protocols declare the predicates that an action and fluent are expected
to define:

\begin{lstlisting}
:- protocol(action_protocol).

	:- public(do/2).
	:- info(do/2, [
		comment is 'True if doing action in ``S1`` results in ``S2``.',
		argnames is ['S1', 'S2']
	]).

	:- public(poss/1).
	:- info(poss/1, [
		comment is 'True if the action is possible in the situation.',
		argnames is ['Situation']
	]).

:- end_protocol.

:- protocol(fluent_protocol).

	:- public(holds/1).
	:- info(holds/1, [
		comment is 'True if the fluent holds in the situation.',
		argnames is ['Situation']
	]).

:- end_protocol.
\end{lstlisting}

Thus any object that is an action or fluent can be found or validated, using the Logtalk built-in
reflection predicates\footnote{The \texttt{conforms\_to\_protocol/2} predicate enumerates both objects and categories that implement a protocol. As we are only interested in objects, we use
the \texttt{current\_object/1} predicate to filter out any categories as these are used only 
to provide common definitions for utility predicates.}. For some strategies attempted without an
interface in OWLSAI, such as when passing the definition context explicitly (as previously
discussed in \myRef{Section}{sec:whitbyarch}), the enumeration of modules requires a hand-coded
alternative to mark the modules, which is fragile and not self-documenting. These predicates are
used to validate or enumerate:
\begin{lstlisting}
is_action(Action) :-
  conforms_to_protocol(Action, action_protocol),
  current_object(Action).

is_fluent(Fluent) :-
  conforms_to_protocol(Fluent, fluent_protocol),
  current_object(Fluent).
\end{lstlisting}
	
Now within the \texttt{sitcalc} object when it is necessary to call a fluent or
action they can be called, even if the argument is a variable, without
depending on the fluents or actions. Here are two extractions from the code
within \texttt{sitcalc} that demonstrate doing so:

\begin{lstlisting}
holds(Fluent, Situation) :-
  is_fluent(Fluent),
  Fluent::holds(Situation).

poss(Action, Situation) :-
  is_action(Action),
  Action::poss(Situation).
\end{lstlisting}

In addition to this, the extraneous code in Golog is not included in
\texttt{SitCalc} such that unused code is not depended upon. Also there is more than
one way to represent a situation in Situation Calculus: either as a history of
actions or as a collection of fluents. Therefore in the publicly available
version\footnote{Comprised of the libraries made available at:
\texttt{https://github.com/PaulBrownMagic/Situations},
\texttt{https://github.com/PaulBrownMagic/Sitcalc}, and
\texttt{https://github.com/PaulBrownMagic/STRIPState}} of the \texttt{SitCalc}
library, the common parts of both
representations are combined into a \texttt{situations} category, with both
representations importing it to ease substitution.

The final detail abstracted from \myRef{Fig}{fig:whitby} is the definition of
action and fluent categories, which import their respective protocols. The
action category defines the \texttt{do/2} predicate and the fluent category
applies tabled resolution to \texttt{holds/2} if available in the backend, which
greatly improves performance of context-dependent queries over long situation terms.

\subsection{Extending SitCalc with Reuseable Libraries}

The two \texttt{OntologyAuthoring} and \texttt{Scaffolding} libraries both
extend \texttt{SitCalc}, but both are also defined in a way that they can be
used as third-party libraries with \texttt{SitCalc} as a dependency.
They extend \texttt{SitCalc} by defining fluents and actions that are
pertinent. \texttt{OntologyAuthoring} includes a fluent to see what triples
hold in the initial situation. Here, \texttt{s0} is a \textit{marker protocol}, allowing easy enumeration of initial situations by using the reflection predicates, and also dependency inversion via a protocol (\texttt{fluent} is a category that implements \texttt{fluent\_protocol}):

\begin{lstlisting}
:- object(initial_assertion(_Subject_, _Predicate_, _Object_),
  imports(fluent)).

  holds(_AnySit) :-
    conforms_to_protocol(S0, s0),
    current_object(S0),
    S0::asserted(_Subject_, _Predicate_, _Object_).

:- end_object.
\end{lstlisting}

\texttt{Scaffolding} includes an action to intervene (here \texttt{action} is a category that implements \texttt{action\_protocol}):

\begin{lstlisting}
:- object(intervene(_Intervention_, _Query_, _Lvl_, _Time_),
  imports(action)).

  poss(Sit) :-
    conforms_to_protocol(Interventions, interventions),
    current_object(Interventions),
    Interventions::intervention(_Intervention_, _Query_),
    sitcalc::holds(_Query_, Sit),
    intervention_level(_Intervention_, _Query_, _Lvl_)::holds(Sit),
    \+ live_intervention(_Intervention_, _Query_, _Lvl_)::holds(Sit).

:- end_object.
\end{lstlisting}

Both objects are \textit{parametric} objects \cite{pmoura11a}. The object parameters (e.g.
\texttt{\_Subject\_} are logic variables shared with all the object predicates.

Due to the implementation of \texttt{SitCalc}, all these fluents and actions
are visible to \texttt{sitcalc} whilst it does not depend on them. However,
these two examples both depend upon some implementation details: some
\texttt{S0::asserted/3} and some \texttt{interventions::intervention/2}. These
dependency issues are solved in the same manner as for \texttt{SitCalc}:
through dependency on a protocol (as illustrated in \myRef{Fig}{fig:whitby}).

Between these two libraries a total of 14 fluent and action terms are
introduced that can be queried via \texttt{SitCalc}. Although these
depend on \texttt{SitCalc}, they do not depend on any application that makes use
of them. Whitby is such an application, by importing these libraries it gains
these 14 fluents and actions, needing only to implement both the \texttt{s0\_protocol} and
\texttt{intervention\_protocol}. In contrast to OWLSAI, the contingent scaffolding
is also not dependent on code that includes ontology authoring, meaning it can
be applied to other activities than ontology authoring.

\section{Conclusion}
Taking a set of rules and applying them to some facts is a typical task in
Prolog. However, the limitations of the module system often result in code that
only handles a fixed set of facts at a time, either imported into the rules module
or loaded into the \texttt{user} special module. But sometimes these rules are
useful to many applications, as is the case with Situation Calculus. When the
rules are to be shared as third-party libraries, any dependency of rules on facts
needs to be inverted to decouple the rules from a particular set of facts. This
dependency inversion allows multiple set of facts to be loaded and used concurrently
(providing an alternative solution for implementing the \textit{many-worlds}
design pattern). Key to this dependency inversion is the concept of
\textit{interface} or \textit{protocol}, supported by Logtalk but absent in
Prolog module systems.

This inversion was achieved in Logtalk by taking inspiration from the
Abstract Factory design pattern and considering how it could be
achieved with \textit{protocols} and \textit{categories}. The final solution
is simpler than the
Abstract Factory design pattern as no dynamic creation of objects is
necessary. Instead, dependency upon a \textit{protocol} and conforming to it is all that
is required. This is an elegant pattern for Logtalk that can be repeated when creating
third-party libraries to reason about definitions in an application without
depending upon them.

The use of \textit{protocols} in this manner results in a \textit{plugin architecture}.
A third-party can ``plug-in'' code to the \texttt{SitCalc} library, or other libraries,
to work with it. This is a very versatile design pattern as it allows an 
application developer, or even third-parties and end-users provided with a plugin
loading interface, to adapt the behaviour of the application to their needs without
editing the core application code. It also leaves the application immune from changes made
elsewhere via the plugin, with the provision they are not malicious, by the drawing of
boundaries in the architecture~\cite{martin:2018}.

This use of \textit{protocols} has focused on their application for dependency inversion due
to the specifics of the Whitby application architecture. It should be noted their use also resulted
in adherence to the Single Responsibility and Open-Closed Principles. Protocols also
have significant contribution to adherence to the Liskov Substitution Principle, making it a
simple matter to swap objects that adhere to the same protocol, as well as the Interface
Segregation Principle by providing explicitly defined interfaces as first class entities.

The refactor from OWLSAI to Whitby decoupled code from OWLSAI that can be reused, which are
published as third-party libraries to satisfy the motivation behind the refactoring. This
has simplified Whitby, where there is less functionality now to maintain, and has enabled
other applications and libraries to use Situation Calculus reasoning while also keeping a
clean architecture. The workarounds that we attempted to compensate for the lack of required
features in the Prolog module systems accumulated and increased the complexity of the application.
Those workarounds are not supported by development tools (especially documenting and diagramming
tools) and raised new issues, thus creating additional burden on developers while not solving the
reusable goals that prompted the refactoring.

By using the language constructs provided by Logtalk to apply SOLID principles in the
refactoring, the Whitby application documentation and diagrams trivially reflect the
actual architecture of the application, further simplifying development
and maintenance. But hand-coded workarounds that try to compensate for missing language
features (in this case: the module system in the original version of the application)
required additional effort to document as they are not visible to developer tools
as first-class constructs. These workarounds must also be repeated in every application
with the impact of their limitations carefully taken into account.

This refactoring has benefited the Whitby application, the Situation Calculus reasoning
is open to extension without modification, which was used to add application specific
fluents and actions as the need arose. Additionally, the separation of responsibilities
has made it easier to navigate and edit the code base. But the primary
benefit is to other applications that wish to make use of the extracted
libraries. Whitby demonstrates how they can be reused.
Bedsit\footnote{\texttt{https://github.com/PaulBrownMagic/BedSit}} is one example of such reuse:
it is an exploratory framework for rapidly prototyping applications using
\texttt{SitCalc} and includes both TicTacToe and ToDo example applications with
a variety of UIs. The first author has also reused \texttt{SitCalc}
and \texttt{OntologyAuthoring} to quickly prototype a proprietary ontology
browser and editor.

As part of the AI4EU\footnote{\texttt{https://www.ai4eu.eu/} Established to build
the first European Artificial Intelligence On-Demand Platform and Ecosystem
with the support of the European Commission under the H2020 programme.} initiative, a
third-party has been provided with Whitby to adapt to a new project in the
domain of robotics planning. Due to Whitby's adherence to the Single Responsibility 
and Liskov Substitution
Principles, which was not possible with OWLSAI, the third-party should need only to make
changes at the periphery of the code base: telling \texttt{kb} to load a different
OWL file, optionally substituting any reasoning rules specific to their domain, substituting \texttt{intervention\_bank} for an object with appropriate
interventions, and substituting the \texttt{Editor} GUI to be appropriate for that
domain. There is still room for improvement, however. For example, they also need
to change a list in \texttt{action\_bank}, which contains the classes used as tabs in the GUI.

Whitby is currently deployed to test the efficacy of the pedagogical techniques implemented.
Should the application prove useful, any remaining architectural issues will be
addressed, although the lesson learned regarding using dependency inversion to decouple
the abstract rules from concrete facts is consistently
applied in all other current software development efforts.

\subsubsection*{Acknowledgements}
The authors gratefully acknowledge the financial support provided: an
EPSRC CASE studentship partially funded by the Defence Science and
Technology Laboratory.  The fourth author is partially funded by the EU AI4EU project (825619) and is a Fellow of the Alan Turing Institute.

\bibliographystyle{acmtrans}
\bibliography{references}


\begin{thebibliography}{00}


\ifx \showCODEN    \undefined \def \showCODEN     #1{\unskip}     \fi
\ifx \showDOI      \undefined \def \showDOI       #1{#1}\fi
\ifx \showISBNx    \undefined \def \showISBNx     #1{\unskip}     \fi
\ifx \showISBNxiii \undefined \def \showISBNxiii  #1{\unskip}     \fi
\ifx \showISSN     \undefined \def \showISSN      #1{\unskip}     \fi
\ifx \showLCCN     \undefined \def \showLCCN      #1{\unskip}     \fi
\ifx \shownote     \undefined \def \shownote      #1{#1}          \fi
\ifx \showarticletitle \undefined \def \showarticletitle #1{#1}   \fi
\ifx \showURL      \undefined \def \showURL       {\relax}        \fi
\providecommand\bibfield[2]{#2}
\providecommand\bibinfo[2]{#2}
\providecommand\natexlab[1]{#1}
\providecommand\showeprint[2][]{arXiv:#2}

\bibitem[\protect\citeauthoryear{Gamma, Helm, Johnson, and Vlissides}{Gamma
  et~al\mbox{.}}{1997}]%
        {gof:1997}
\bibfield{author}{\bibinfo{person}{Erich Gamma}, \bibinfo{person}{Richard
  Helm}, \bibinfo{person}{Ralph Johnson}, {and} \bibinfo{person}{John
  Vlissides}.} \bibinfo{year}{1997}\natexlab{}.
\newblock \bibinfo{booktitle}{{\em Design Patterns: Elements of Reusable
  Object-Oriented Software}}.
\newblock \bibinfo{publisher}{Addison Wesley}, \bibinfo{address}{Reading,
  Massachusetts}.
\newblock


\bibitem[\protect\citeauthoryear{Gruber}{Gruber}{1995}]%
        {gruber:1995}
\bibfield{author}{\bibinfo{person}{Thomas~R. Gruber}.}
  \bibinfo{year}{1995}\natexlab{}.
\newblock \showarticletitle{Toward principles for the design of ontologies used
  for knowledge sharing?}
\newblock \bibinfo{journal}{{\em International Journal of Human-Computer
  Studies\/}} \bibinfo{volume}{43}, \bibinfo{number}{5-6}
  (\bibinfo{year}{1995}), \bibinfo{pages}{907--928}.
\newblock


\bibitem[\protect\citeauthoryear{ISO/IEC}{ISO/IEC}{1995}]%
        {isoiec95}
\bibfield{author}{\bibinfo{person}{ISO/IEC}.} \bibinfo{year}{1995}\natexlab{}.
\newblock \bibinfo{booktitle}{{\em International Standard ISO/IEC 13211-1
  Information Technology --- Programming Languages --- Prolog --- Part I:
  General core}}.
\newblock \bibinfo{publisher}{ISO/IEC}.
\newblock


\bibitem[\protect\citeauthoryear{ISO/IEC}{ISO/IEC}{2000}]%
        {isoiec00}
\bibfield{author}{\bibinfo{person}{ISO/IEC}.} \bibinfo{year}{2000}\natexlab{}.
\newblock \bibinfo{booktitle}{{\em International Standard ISO/IEC 13211-2
  Information Technology --- Programming Languages --- Prolog --- Part II:
  Modules}}.
\newblock \bibinfo{publisher}{ISO/IEC}.
\newblock


\bibitem[\protect\citeauthoryear{Martin}{Martin}{2018}]%
        {martin:2018}
\bibfield{author}{\bibinfo{person}{Robert~C Martin}.}
  \bibinfo{year}{2018}\natexlab{}.
\newblock \bibinfo{booktitle}{{\em Clean Architecture: A Craftman's Guide to
  Software Structure and Design}}.
\newblock \bibinfo{publisher}{Prentice Hall}, \bibinfo{address}{Hudson, New
  Jersey}.
\newblock


\bibitem[\protect\citeauthoryear{Moura}{Moura}{2011}]%
        {pmoura11a}
\bibfield{author}{\bibinfo{person}{Paulo Moura}.}
  \bibinfo{year}{2011}\natexlab{}.
\newblock \showarticletitle{Programming Patterns for Logtalk Parametric
  Objects}.
\newblock In \bibinfo{booktitle}{{\em Applications of Declarative Programming
  and Knowledge Management}}, \bibfield{editor}{\bibinfo{person}{Salvador
  Abreu} {and} \bibinfo{person}{Dietmar Seipel}} (Eds.).
  \bibinfo{series}{Lecture Notes in Artificial Intelligence},
  Vol.~\bibinfo{volume}{6547}. \bibinfo{publisher}{Springer-Verlag},
  \bibinfo{address}{Berlin Heidelberg}, \bibinfo{pages}{52--69}.
\newblock


\bibitem[\protect\citeauthoryear{Moura}{Moura}{2021}]%
        {TheLogtalkHandbook}
\bibfield{author}{\bibinfo{person}{Paulo Moura}.}
  \bibinfo{year}{2021}\natexlab{}.
\newblock \bibinfo{booktitle}{{\em {The Logtalk Handbook}\/}
  (\bibinfo{edition}{{Release 3.46.0}} ed.)}.
\newblock


\bibitem[\protect\citeauthoryear{Reiter}{Reiter}{2001}]%
        {reiter:2001}
\bibfield{author}{\bibinfo{person}{Raymond Reiter}.}
  \bibinfo{year}{2001}\natexlab{}.
\newblock \bibinfo{booktitle}{{\em Knowledge in Action}}.
\newblock \bibinfo{publisher}{The {MIT} Press}, \bibinfo{address}{Cambridge,
  Massachusetts}.
\newblock


\bibitem[\protect\citeauthoryear{Wood, Bruner, and Ross}{Wood
  et~al\mbox{.}}{1976}]%
        {wood:1976}
\bibfield{author}{\bibinfo{person}{David Wood}, \bibinfo{person}{Jerome~S.
  Bruner}, {and} \bibinfo{person}{Gail Ross}.} \bibinfo{year}{1976}\natexlab{}.
\newblock \showarticletitle{The Role of Tutoring in Problem Solving}.
\newblock \bibinfo{journal}{{\em Journal of Child Psychology and Psychiatry\/}}
  \bibinfo{volume}{17}, \bibinfo{number}{2} (\bibinfo{year}{1976}),
  \bibinfo{pages}{89--100}.
\newblock


\end{thebibliography}

\label{lastpage}
\end{document}